\documentclass[10pt, twocolumn, secnumarabic, amssymb, nobibnotes, footinbib, showpacs, aps, pra, superscriptaddress]{revtex4-1}

\usepackage{graphicx}
\usepackage{dcolumn}
\usepackage{amsmath}
\usepackage{natbib}
\usepackage{color}
\usepackage{epstopdf}

\begin{document}

%%%%%%%%%%%%%%%%%%%%%%%%%%%%%%%%%%%%%%%%%%%%%%%%%%%%%%%%%%%%%%%%%%%%

\title{Three-dimensional visualization of a qutrit}

\author{Pawe\l{} Kurzy\'nski}   \email{pawel.kurzynski@amu.edu.pl}   \affiliation{Faculty of Physics, Adam Mickiewicz University, Umultowska 85, 61-614 Pozna\'n, Poland}
 \affiliation{Centre for Quantum Technologies, National University of Singapore, 3 Science Drive 2, 117543 Singapore, Singapore}

\author{Adrian Ko\l{}odziejski}
\author{Wies{\l}aw Laskowski}
\affiliation{Institute of Theoretical Physics and Astrophysics, University of Gda\'nsk, 80-952 Gda\'nsk, Poland}

\author{Marcin Markiewicz}     
\affiliation{Institute of Physics, Jagiellonian University, ul. \L{}ojasiewicza 11,
30-348 Krak\'ow, Poland}

\date{\today}

%%%%%%%%%%%%%%%%%%%%%%%%%%%%%%%%%%%%%%%%%%%%%%%%%%%%%%%%%%%%%%%%%%%%

\begin{abstract}
We present a surprisingly simple three-dimensional Bloch sphere representation of a qutrit, i.e., a single three-level quantum system. We start with a symmetric state of a two-qubit system and relate it to the spin-1 representation. Using this representation we associate each qutrit state with a three-dimensional vector $\mathbf{a}$ and a metric tensor $\mathbf{\hat\Gamma}$ which satisfy $\mathbf{a}\cdot \mathbf{\hat\Gamma} \cdot \mathbf{a}\leq 1$. This resembles the well known condition for qubit Bloch vectors in which case $\mathbf{\hat\Gamma}=\openone$. In our case the vector $\mathbf{a}$ corresponds to spin-1 polarization, whereas the tensor $\mathbf{\hat\Gamma}$ is a function of polarization uncertainties. Alternatively, $\mathbf{a}$ is a local Bloch vector of a symmetric two-qubit state and $\mathbf{\hat\Gamma}$ is a function of the corresponding correlation tensor.
\end{abstract}


\maketitle

%%%%%%%%%%%%%%%%%%%%%%%%%%%%%%%%%%%%%%%%%%%%%%%%%%%%%%%%%%%%%%%%%%%%

\section{Introduction} 

The popularity of a Bloch sphere representation of a qubit stems from its simplicity. The state of a qubit can be represented by a simple three-dimensional vector $\mathbf{a}$ and any vector obeying $|\mathbf{a}| \leq 1$ corresponds to some valid qubit state. The other way around, any qubit state can be associated with a unique vector whose length is less than one. This property offers a simple way of representing both, quantum states and quantum dynamics of a single-qubit system. 

There were some attempts to generalize the Bloch sphere representation to higher-dimensional systems \cite{Kimura1, Bertlmann, Kimura2, Mendas, KZ, JS, GOYAL, ASADIAN}. Because $d$-level quantum systems are described by $d^2-1$ parameters one cannot represent their states solely by single three-dimensional vectors (apart from the obvious case of $d=2$). The usual method is to consider vectors in more than three dimensions, however such representation cannot be easily grasped by our intuition. Moreover, unlike in $d=2$ case, a set of constraints on such multi-dimensional vectors does not posses rotational symmetry. Alternatively, one can use more than one three-dimensional vector to represent a state, however the constraints on these vectors are also not intuitive \cite{Pawel}. It is therefore important to look for new simple methods to graphically represent states of higher-dimensional quantum systems.

Here we present a three-dimensional graphical representation of a qutrit ($d=3$) using a single vector $\mathbf{a}$ and a metric tensor $\mathbf{\hat\Gamma}$. The constraint on a valid density matrix resembles the one for qubit, i.e., $\mathbf{a}\cdot \mathbf{\hat\Gamma} \cdot \mathbf{a}\leq 1$, where in qubit case the metric tensor is just identity. Finally, we discuss how to describe a purity of states and a dynamics using our method.    

%%%%%%%%%%%%%%%%%%%%%%%%%%%%%%%%%%%%%%%%%%%%%%%%%%%%%%%%%%%%%%%%%%%%

\section{Qutrit representation} 

Let us start with two representations of a qutrit. First, we represent it as a symmetric state of two qubits. An arbitrary two-qubit density matrix can be written in the following form:
\begin{eqnarray} 
\rho&=&
\frac{1}{4}\left(\openone \otimes \openone + \sum_{j=x,y,z}\left(a_j \sigma_j \otimes \openone + b_j \openone\otimes \sigma_j\right)\right. \nonumber \\ &+& \left. \sum_{j,k=x,y,z}T_{jk} \sigma_j \otimes \sigma_k \right),
\end{eqnarray}
where $\mathbf{a}=(a_x,a_y,a_z)$ and $\mathbf{b}=(b_x,b_y,b_z)$ are local Bloch vectors and $T_{jk}$ are elements of a correlation tensor  $\mathbf{\hat T}$.

In case of symmetric two-qubit states $\mathbf{a}=\mathbf{b}$ and $T_{jk}=T_{kj}$. Moreover, it can be easily shown that, since the symmetric state is orthogonal to the singlet state $|\psi_-\rangle\langle \psi_-|$, one has $\text{Tr}(\mathbf{\hat T})=1$.

Next, we identify the symmetric two-qubit operators with the spin-1 matrices. We get
\begin{eqnarray}
S_j &=& \frac{\sigma_{j}\otimes \openone + \openone \otimes \sigma_{j}}{2}, \nonumber \\
S_j^2 &=& \frac{\openone\otimes \openone + \sigma_{j} \otimes \sigma_{j}}{2}, \label{A} \\
A_j &=& S_k S_l + S_l S_k \\
    &=& \frac{\sigma_k \otimes \sigma_l + \sigma_l \otimes \sigma_k}{2} ~~(j \neq k \neq l). \nonumber 
\end{eqnarray}
From this we have
\begin{eqnarray}
\langle S_j \rangle &=& a_j, \\
\langle S_j^2 \rangle &=& \frac{1 + T_{jj}}{2}, \\
\langle A_j \rangle &=& T_{kl} = T_{lk} = q_j ~~(j \neq k \neq l).
\end{eqnarray}
It is also convenient to introduce
\begin{equation}
1-\langle S_j^2 \rangle = \frac{1 - T_{jj}}{2}= \omega_j.
\end{equation}
that satisfies
\begin{equation}
\sum_{j=x,y,z} \omega_j = 1.
\end{equation}

Let us now introduce an alternative representation of spin-1 matrices. An interesting property is that in case of spin-1 $S_x^2$, $S_y^2$ and $S_z^2$ mutually commute hence they can be represented as diagonal matrices
\begin{equation}
S_x^2=\begin{pmatrix} 0 & 0 & 0 \\ 0 & 1 & 0 \\ 0 & 0 & 1 \end{pmatrix},~~S_y^2=\begin{pmatrix} 1 & 0 & 0 \\ 0 & 0 & 0 \\ 0 & 0 & 1 \end{pmatrix}, S_z^2=\begin{pmatrix} 1 & 0 & 0 \\ 0 & 1 & 0 \\ 0 & 0 & 0 \end{pmatrix}.\label{CC}
\end{equation}
The corresponding $S_x$, $S_y$ and $S_z$ matrices are of the following form
\begin{equation}
S_x=\begin{pmatrix} 0 & 0 & 0 \\ 0 & 0 & -i \\ 0 & i & 0 \end{pmatrix},
~~S_y=\begin{pmatrix} 0 & 0 & i \\ 0 & 0 & 0 \\ -i & 0 & 0 \end{pmatrix}, 
S_z=\begin{pmatrix} 0 & -i & 0 \\ i & 0 & 0 \\ 0 & 0 & 0 \end{pmatrix}.
\label{BB}
\end{equation}

Since the $4\times 4$ matrices (\ref{A}) can be represented in a block form $3 \oplus 1$, we can express a symmetric state of two qubits using the above $3 \times 3$ spin-1 matrices. In order to span the whole operator space one should also use additional three matrices $A_x$, $A_y$ and $A_z$, see Eq. (\ref{A}),
\begin{equation}
\rho= \sum_{j=x,y,z} \left( \omega_j(\openone - S_j^2) + \frac{a_j S_j + q_j A_j}{2} \right).
\end{equation}
In the matrix form we get
\begin{equation}
\rho=\begin{pmatrix}
\omega_x & \frac{-i a_z-q_z}{2}& \frac{i a_y-q_y}{2}\\
\frac{i a_z-q_z}{2}& \omega_y & \frac{-i a_x-q_x}{2}\\
\frac{-i a_y-q_y}{2} & \frac{i a_x-q_x}{2} & \omega_z
\end{pmatrix}.
\end{equation}
We recall that $q_j=T_{kl}=T_{lk}$ and $\omega_j=\frac{1-T_{jj}}{2}$, which gives
\begin{equation}
\rho= \frac{1}{2}\begin{pmatrix}
1-T_{xx} & -i a_z-T_{xy}& i a_y-T_{xz}\\
i a_z-T_{yx}& 1-T_{yy} & -i a_x-T_{yz}\\
-i a_y-T_{zx} & i a_x-T_{zy} & 1-T_{zz}
\end{pmatrix}.
\end{equation}
Therefore, the correlation tensor can be easily obtained from the density matrix
\begin{equation}
\hat{\mathbf{T}}=\openone - 2\Re(\rho),
\end{equation}
where $\Re(\rho)$ denotes the real part of $\rho$.

Next, for the sake of clarity of presentation, only the case of $q_j = 0$ for all $j$ is shown (the correlation tensor is diagonal).  
Thus, we can write
\begin{equation}\label{diag}
\rho=\begin{pmatrix}
\omega_x & \frac{-i a_z}{2} & \frac{i a_y}{2} \\
\frac{i a_z}{2} & \omega_y & \frac{-i a_x}{2} \\
\frac{-i a_y}{2} & \frac{i a_x}{2} & \omega_z
\end{pmatrix}.
\end{equation}

%%%%%%%%%%%%%%%%%%%%%%%%%%%%%%%%%%%%%%%%%%%%%%%%%%%%%%%%%%%%%%%%%%%%

\section{Conditions for non-negativity of the density matrix} 

In this section we base on a previous work of one of us \cite{Pawel} and present a set of necessary and sufficient conditions for non-negativity of a qutrit density matrix. A hermitian matrix is positive-semidefinite if determinants of all principal minors of this matrix are non-negative \cite{Horn}. A principal minor is a determinant of a sub-matrix defined upon the diagonal of the original matrix. Therefore, we need to check the non-negativity of all diagonal elements, determinants of all $2\times 2$ principal sub-matrices and finally the determinant of $\rho$.

In general $|a_j|\leq 1$ and $0 \leq \omega_j \leq 1$. The conditions resulting from non-negativity of diagonal elements (which are less than 1) are 
\begin{equation}\label{c1}
1 \geq \omega_j \geq 0,
\end{equation}
or in terms of the correlation tensor
\begin{equation}
1 \geq T_{jj} \geq -1.
\end{equation}

Next, non-negativity of $2\times 2$ principal minors leads to the following conditions 
\begin{equation}\label{c2}
4\omega_j \omega_k \geq a_l^2,
\end{equation}
or
\begin{equation}
(1-T_{jj})(1-T_{kk}) \geq a_l^2,
\end{equation}
where $j\neq k \neq l$ (from now on this will also be true in any other case in which these three indices appear together, unless otherwise stated).

Finally, the non-negativity of the determinant of $\rho$ gives:
\begin{equation}\label{f1}
4\omega_x \omega_y \omega_z \geq \omega_x a_x^2 + \omega_y a_y^2 + \omega_z a_z^2.
\end{equation}
The above can be rewritten as
\begin{equation}\label{c3}
1 \geq  \mathbf{a}\cdot \mathbf{\hat\Gamma} \cdot \mathbf{a},
\end{equation}
where $\mathbf{\hat\Gamma}$ can be interpreted as a metric tensor of the form
\begin{equation}\label{Gamma}
\mathbf{\hat\Gamma} \equiv \frac{\openone- \mathbf{\hat T}}{\text{det}(\openone- \mathbf{\hat T})} = \frac{\Re(\rho)}{4~\text{det}\left(\Re(\rho)\right)}.
\end{equation}

If one calculates the determinant of $\rho$ without assuming that $\mathbf{\hat T}$ is diagonal, one gets a slightly more complicated formula than (\ref{f1}). However, what is interesting, is that this formula can be also rewritten in the form (\ref{c3}), but this time the metric tensor is not diagonal. In addition, (\ref{c3}) implies (\ref{c2}). Therefore, we can relax the assumption about the diagonal form of $\mathbf{\hat T}$.

%%%%%%%%%%%%%%%%%%%%%%%%%%%%%%%%%%%%%%%%%%%%%%%%%%%%%%%%%%%%%%%%%%%%

\section{Properties of the metric tensor} 

Tensor $\mathbf{\hat\Gamma}$ is a function of the correlation tensor $\mathbf{\hat T}$. The trace of the correlation tensor is 1 and its eigenvalues $\lambda_j$ are bounded by $\pm 1$, therefore, according to the definition (\ref{Gamma}), $\mathbf{\hat\Gamma}$ is positive-semidefinite. The index $j$ labels the eigenbasis directions ($j=u,v,w$). We arrange the eigenvalues of $\mathbf{\hat T}$ in the following order $1\geq \lambda_u \geq \lambda_v \geq \lambda_w \geq -1$

One can easily verify that at most one eigenvalue of $\mathbf{\hat T}$ can be negative and that if $\lambda_w < 0$, then $\lambda_u \geq \lambda_v \geq |\lambda_w|$. Eigenvalues of $\mathbf{\hat\Gamma}$ are nonnegative and can be expressed as
\begin{equation}
\gamma_j=\frac{1}{(1-\lambda_k)(1-\lambda_l)}.
\end{equation} 
Because of the eigenvalue arrangement we can write the following bounds on $\gamma_j$ 
\begin{equation}
\gamma_j\geq \frac{1}{1-\lambda_w^2}\geq 1.
\end{equation}

%%%%%%%%%%%%%%%%%%%%%%%%%%%%%%%%%%%%%%%%%%%%%%%%%%%%%%%%%%%%%%%%%%%%

\section{Graphical representation} 

Up to now we know that a qutrit can be represented by a three-dimensional vector and a metric tensor. A vector is described by three parameters and a tensor is described by five -- three of them correspond to Euler angles defining its principal directions and the remaining two define its eigenvalues (two not three, because of the constraint $\text{Tr}(\mathbf{\hat T})=1$). There are eight parameters in total which guarantees a complete description of a qutrit.  Our next goal is to show how to graphically represent them in a three-dimensional space.

\begin{figure}[t]
    \begin{center}
     %Zmieni?em tutaj z eps na pdf bo nie mog?em zrobi? rysunków w eps, Pawe?
	\includegraphics[width=0.5\columnwidth]{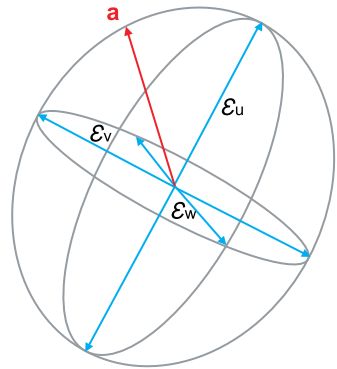}
        \caption{Three-dimensional graphical representation of a qutrit. The state is described by an ellipsoid (5 parameters) which contains a vector $\mathbf{a}$ (3 parameters).}
        \label{fig1}
    \end{center}
\end{figure}

For a fixed $\mathbf{\hat\Gamma}$ the constraint (\ref{c3}) implies that a set of allowed vectors $\mathbf{a}$ is given by an ellipsoid whose shape and orientation in space, i.e., direction of semi-principal axes, is determined by the eigenvalues and the eigenvectors of $\mathbf{\hat T}$. The lengths of the semi-principal axes are given by 
\begin{equation}\label{eps}
\varepsilon_j=1/\sqrt{\gamma_j}=\sqrt{(1-\lambda_k)(1-\lambda_l)} \leq \sqrt{1-\lambda_w^2}, 
\end{equation}
therefore $1 \geq \varepsilon_j \geq 0$. This resembles the Bloch sphere of a qubit, for which all semi-principal axes are of length one for every qubit state. In case of a qutrit semi-principal axes can have different lengths and can point in various directions in space -- see Fig. \ref{fig1}.

Using the above notation one can reformulate the constraints (\ref{c2}) as
\begin{equation}\label{cs2}
\varepsilon_j \geq |a_j|
\end{equation}

Next, we consider three cases. (i) $1> \lambda_u \geq \lambda_v \geq \lambda_u$. In this case all $\varepsilon_j > 0$ and the ellipsoid is a three-dimensional object. (ii) $\lambda_u = 1$, but $1 > \lambda_v=-\lambda_w$. In this case $\varepsilon_u >0$, but $\varepsilon_v = \varepsilon_w = 0$ and the ellipsoid becomes a one-dimensional object -- a line segment directed along $\mathbf{u}$. In this case the vector $\mathbf{a}$ must lie on this segment. (iii) $\lambda_u = \lambda_v = -\lambda_w= 1$. In this case all $\varepsilon_j=0$ and the ellipsoid becomes a single point at the origin. In this case $\mathbf{a}=\mathbf{0}$. 

Before we proceed, we need to mention one important feature. In case (ii) the system is visualized by a line segment and a vector lying along it. This segment reveals information about only one principal direction of the metric tensor. In case (iii) we have a single point which does not give any information about the orientation of the metric tensor. In such cases one does not have enough data to uniquely reconstruct the corresponding density matrix. It is therefore necessary to add some extra information to our visualisation. 

\begin{figure}[t]
    \centering
	\includegraphics[width=0.95\columnwidth]{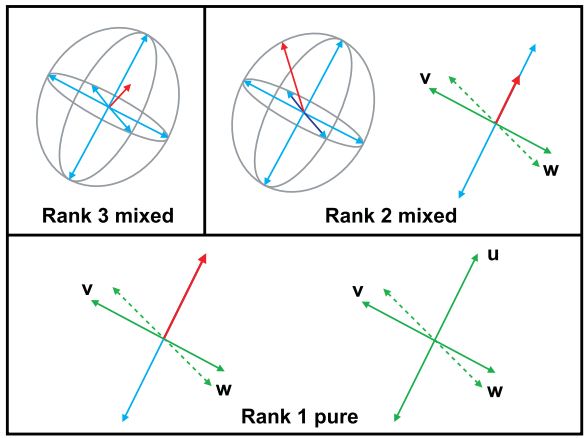}
        \caption{Graphical representation of mixed and pure states. Red arrows denote Bloch vectors $\mathbf{a}$. Blue (green) rays denote (vanishing) semi-principal axes of the ellipsoid. Dashed green rays denote semi-principal axes of the ellipsoid corresponding to the eigenvectors of the correlation tensor with the eigenvalue $-1$.}
        \label{fig2}
\end{figure}

Every time the two semi-principal axes of the ellipsoid vanish (case (ii)) we add two rays to denote the principle directions of the correlation/metric tensor -- the solid ray corresponding to $\lambda_v$ and the dashed one corresponding to $\lambda_w$ ($\lambda_v > \lambda_w$). In case (iii), in which all semi-principal axes vanish, we add a third solid ray corresponding to $\lambda_u$ -- see Fig. \ref{fig2}.

Note that there has been proposed an alternative geometrical construction based on two-qubit states, which also leads to a similar ellipsoid-like picture \cite{JEVTIC}, which however depicted completely different features.

%%%%%%%%%%%%%%%%%%%%%%%%%%%%%%%%%%%%%%%%%%%%%%%%%%%%%%%%%%%%%%%%%%%%

\section{Pure and mixed states} 

One may expect that since pure qubit states correspond to vectors lying on a surface of the Bloch sphere, pure qutrit states will correspond to vectors lying on a surface of the ellipsoid. However, one has to remember that a set of qutrit mixed states has a richer structure than the one for qubits. The most important difference is that for qubits a density matrix of a mixed state is always of rank two. For qutrits such states can be either rank two or rank three. Here, we will show that if the vector $\mathbf{a}$ lies inside the three-dimensional ellipsoid then the corresponding density matrix is of rank three. On the other hand, the state can be pure only if the corresponding ellipsoid is a point or a one-dimensional line segment with $|\mathbf{a}|=\varepsilon_u$. Finally, the state is of rank 2 if $\mathbf{a}$ lies on the surface of the three-dimensional ellipsoid, or  the ellipsoid becomes a line segment and $|\mathbf{a}|<\varepsilon_u$ -- see Fig. \ref{fig2}. 

For simplicity we come back to the basis in which $\mathbf{\hat T}$ and $\mathbf{\hat \Gamma}$ are diagonal, so the state can be represented by Eq. (\ref{diag}). Our main tools are the interlacing theorem \cite{Horn} and the fact that density matrices have non-negative eigenvalues. The interlacing theorem states that eigenvalues of any principal submatrix are interlaced between the eigenvalues of the original hermitian matrix. In particular, if $\lambda_3 \geq \lambda_2 \geq \lambda_1$ are eigenvalues of a $3\times 3$ hermitian matrix and $\lambda'_2 \geq \lambda'_1$ are eigenvalues of one of its $2\times 2$ principal  submatrices, then the following holds
\begin{equation}
\lambda_3 \geq \lambda'_2\geq  \lambda_2 \geq \lambda'_1 \geq \lambda_1.
\end{equation}

If the determinant of $\rho$ is nonzero then the determinants of all $2\times 2$ principal submatrices are also nonzero. In this case conditions (\ref{c3}) and (\ref{cs2}) are strict inequalities and the state is of rank 3. This implies that the ellipsoid is a three-dimensional object and that $\mathbf{a}$ lies inside it. 

If determinants of all $2\times 2$ principal submatrices are nonzero, but the determinant of the $3\times 3$ density matrix is zero, then for each principal submatrix one has $\lambda'_2 \geq \lambda'_1 > 0$, thus $\lambda_3\geq \lambda_2 > \lambda_1=0$ and the qutrit state is of rank two. This happens when condition (\ref{c3}) becomes an equality, i.e., the vector $\mathbf{a}$ lies on the surface of the ellipsoid, but conditions (\ref{cs2}) are strict inequalities. These conditions can be strict inequalities only if $\varepsilon_j > 0$ for all $j$, but this implies that the ellipsoid is a three-dimensional object. 

Alternatively, rank 2 states can arise if the determinant of the $3\times 3$ density matrix and the determinants of two $2\times 2$ principal submatrices are zero, but the determinant of the remaining $2\times 2$ principal submatrix is nonzero. This implies that condition (\ref{c3}) and two of three conditions (\ref{cs2}) are equalities. The remaining condition (\ref{cs2}) is a strict inequality $\varepsilon_u > |a_u|$. In this case the vector $\mathbf{a}$ is shorter than the one-dimensional line segment along which it is lying.  

Finally, the density matrix is of rank one, i.e., a pure state, if only one of its eigenvalues is nonzero. In this case the determinant of $\rho$ and the determinants of all principal submatrices are zero. In other words, the condition (\ref{c3}) and all conditions (\ref{c2}) are equalities. However, this can only happen if the left-hand side of two or three conditions (\ref{c2}) are zero. Otherwise one would violate (\ref{c3}). In this case the ellipsoid is either a line segment or a point, depending on whether the left-hand sides of two or three conditions (\ref{c2}) were zero, respectively. Moreover, if it is a line segment then $|\mathbf{a}|=\varepsilon_u$.

%%%%%%%%%%%%%%%%%%%%%%%%%%%%%%%%%%%%%%%%%%%%%%%%%%%%%%%%%%%%%%%%%%%%

\section{Orthogonality}

Next, we are going to discuss how orthogonality of pure states can be visualized in our representation. We first recall that pure states can be represented as normalized vectors
\begin{equation}
|\psi\rangle = \begin{pmatrix} \alpha \\ \beta \\ \gamma \end{pmatrix},
\end{equation}
where $\alpha \in \mathcal{R}$, $\beta,\gamma \in \mathcal{C}$ and $\alpha^2 + |\beta|^2+|\gamma|^2=1$.  Alternatively, we can represent them as 
\begin{equation}
|\psi\rangle = \mathbf{r} + i \mathbf{k} = \begin{pmatrix} x \\ y \\ z \end{pmatrix} + i \begin{pmatrix} 0 \\ a \\ b \end{pmatrix},
\end{equation}
where $\mathbf{r}$ and $\mathbf{k}$ are real vectors.

\subsection{Real pure states}

In case $|\psi\rangle$ is real, i.e., $\mathbf{k} = \mathbf{0 }$, the Bloch vector vanishes (this is because the density matrix is real and the elements of $\rho$ corresponding to the coordinates of the Bloch vector are imaginary) and pure states are represented by green dashed rays -- corresponding to the eigenvector of the correlation tensor with the eigenvalue $-1$  (see the second case in Fig. \ref{fig2}). The orientation of the green dashed ray in a three-dimensional space is exactly the same as the orientation of $\mathbf{r}$. This is because the space of pure real qutrit states is isomorphic to the 3D Euclidean space. Therefore, two real state vectors ($|\psi\rangle = \mathbf{r}$ and $|\psi'\rangle = \mathbf{r}'$) are orthogonal if the corresponding dashed green rays are orthogonal. 

\subsection{Complex pure states}

In case $|\psi\rangle$ is complex the pure state is represented by both, Bloch vector and green rays. The density matrix corresponding to this state can be expressed in terms of elements of $\mathbf{r}$ and $\mathbf{k}$
\begin{equation}
\rho=\begin{pmatrix}
x^2 & xy -ixa & xz-ixb\\
yx + ixa & y^2 + a^2 & yz +ab -iyb +iaz\\
zx + ixb &  zy + ba + iyb -iaz & z^2 + b^2
\end{pmatrix}.
\end{equation}
From this we get that the Bloch vector is of the form
\begin{equation}
\mathbf{a}=2\begin{pmatrix} yb-az \\ -xb \\ xa \end{pmatrix}.
\end{equation}
Interestingly, we can express it as a cross product  
\begin{equation}
\mathbf{a}=2\mathbf{r}\times \mathbf{k},
\end{equation}
therefore $\mathbf{a}$ is orthogonal to both, $\mathbf{r}$ and $\mathbf{k}$.

In addition, note that 
\begin{equation}
\Re(\rho)\cdot \mathbf{a}= 0,
\end{equation}
therefore $\mathbf{a}$ is an eigenvector of $\openone - \hat{\mathbf{T}}$ corresponding to the eigenvalue $0$. This means that the eigenvalues of $\hat{\mathbf{T}}$ are $1$, $\lambda$ and $-\lambda$, where $0\leq \lambda \leq 1$ and $\mathbf{a}$ is the eigenvector of $\hat{\mathbf{T}}$ corresponding to the eigenvalue $1$. Such pure states are represented by the set of green rays indicating eigenvectors of $\hat{\mathbf{T}}$ and the red Bloch vector lying along direction corresponding to the greatest eigenvalue of $\hat{\mathbf{T}}$ (see the first case in Fig. \ref{fig2}).

Next, we consider orthogonality relation. First, let us discuss the case of a complex state $|\psi\rangle = \mathbf{r} + i \mathbf{k}$ and a real state $|\psi'\rangle = \mathbf{r}'$. These states are orthogonal if $ \mathbf{r}'$ is orthogonal to both, $ \mathbf{r}$ and $ \mathbf{k}$. This means that $ \mathbf{r}'$ lies along $ \mathbf{r} \times  \mathbf{k}$, i.e., the dashed ray of the real state lies along the red Bloch vector of the complex state.

Finally, let us consider orthogonality between two complex states $|\psi\rangle = \mathbf{r} + i \mathbf{k}$ and $|\psi'\rangle = \mathbf{r}' + i \mathbf{k}'$. These two states are orthogonal iff
\begin{equation}
\mathbf{r}\cdot \mathbf{r}' = -\mathbf{k}\cdot \mathbf{k'} ~~\wedge~~  \mathbf{r}\cdot \mathbf{k}' = \mathbf{r}'\cdot \mathbf{k}.
\end{equation}
Without loosing generality we consider a state 
\begin{equation}
|\psi\rangle = \begin{pmatrix} \cos\theta \\ i\sin\theta \\ 0 \end{pmatrix},
\end{equation}
where $\theta \in [0,\pi/2]$. The corresponding Bloch vector is of the form
\begin{equation}
\mathbf{a} = \begin{pmatrix} 0 \\ 0 \\ 2\cos\theta\sin\theta  \end{pmatrix}.
\end{equation}

Any pure state that is orthogonal to $|\psi\rangle$ is given by
\begin{equation}
|\psi'\rangle = \begin{pmatrix} \sin\theta \cos\varphi \\ -i\cos\theta\cos\varphi \\ e^{i\chi}\sin\varphi \end{pmatrix}.
\end{equation}
The corresponding Bloch vector yields
\begin{equation}
\mathbf{a}' = 2\begin{pmatrix} \cos \varphi \sin \varphi \cos \theta \cos\chi \\  \cos \varphi \sin \varphi \sin \theta \sin\chi  \\ -\cos^2 \varphi \cos\theta\sin\theta  \end{pmatrix}.
\end{equation}

\begin{figure}[t]
    \begin{center}	\includegraphics[scale=0.5]{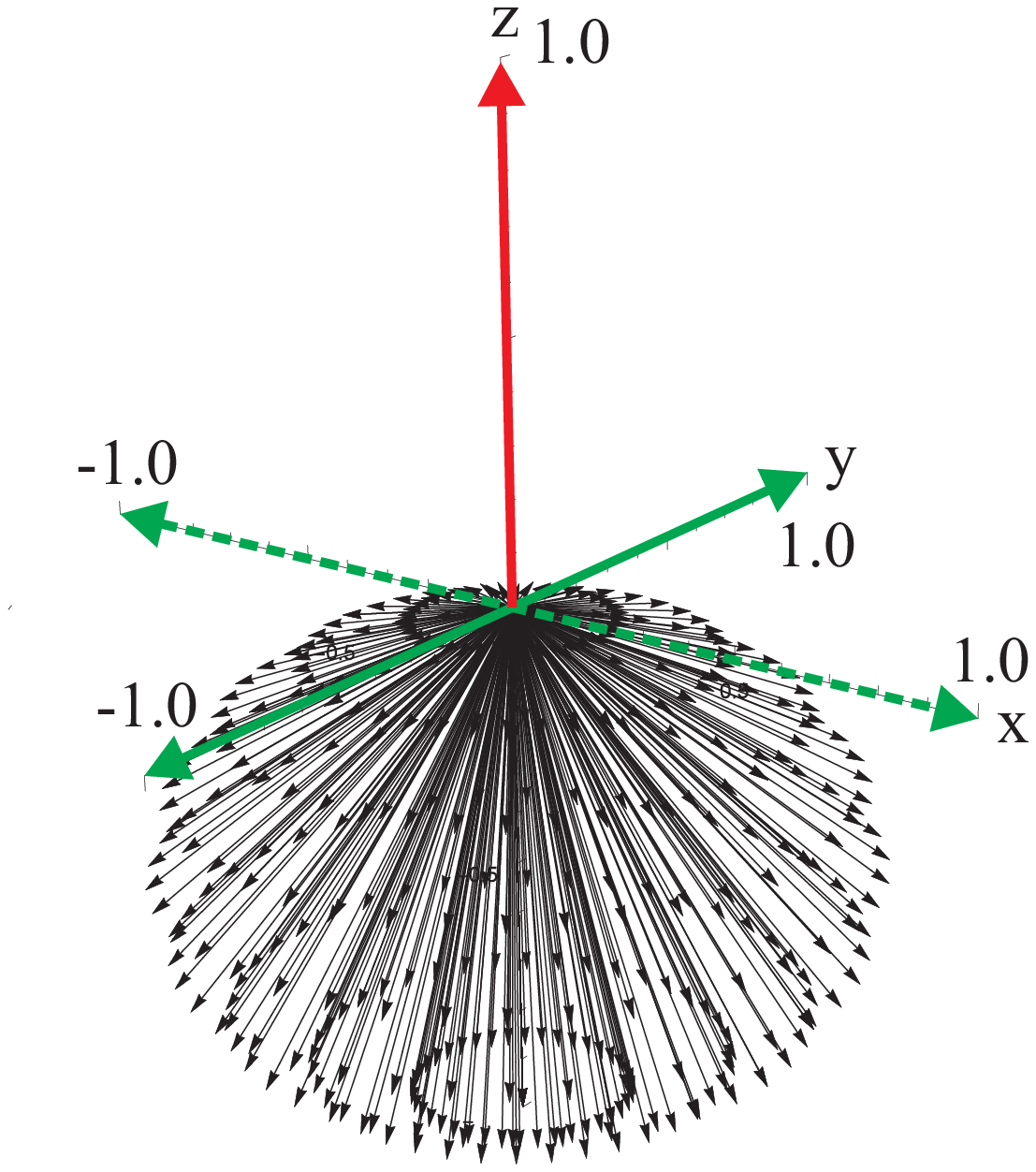}
	\vspace{-3mm}
        \caption{Graphical representation of a pure state $|\psi\rangle=1/\sqrt{2}(|0\rangle - i |1\rangle)$. Black arrows denote Bloch vectors of states that are orthogonal to $|\psi\rangle$.}
        \label{fig4}
    \end{center}
\end{figure}

For a fixed $\theta$ the state $|\psi'\rangle$ and the Bloch vector $\mathbf{a}'$ are described by two parameters $\varphi$ and $\chi$. It is not immediately visible what are the geometrical properties of two complex orthogonal states. We visualize them in Fig. \ref{fig4}, where we provide sets of Bloch vectors corresponding to the states that are orthogonal to $|\psi\rangle$. One can see that a prerequisite for orthogonality of $|\psi\rangle$ and $|\psi'\rangle$ is that $\mathbf{a} \cdot \mathbf{a}' \leq 0$.

%%%%%%%%%%%%%%%%%%%%%%%%%%%%%%%%%%%%%%%%%%%%%%%%%%%%%%%%%%%%%%%%%%%%

\section{Examples}

In this section we present visualisation of two examples of different classes of qutrit states 

\subsection{Mutually unbiased bases}

\begin{figure}[t]
    \begin{center}
	\includegraphics[width=0.98\columnwidth]{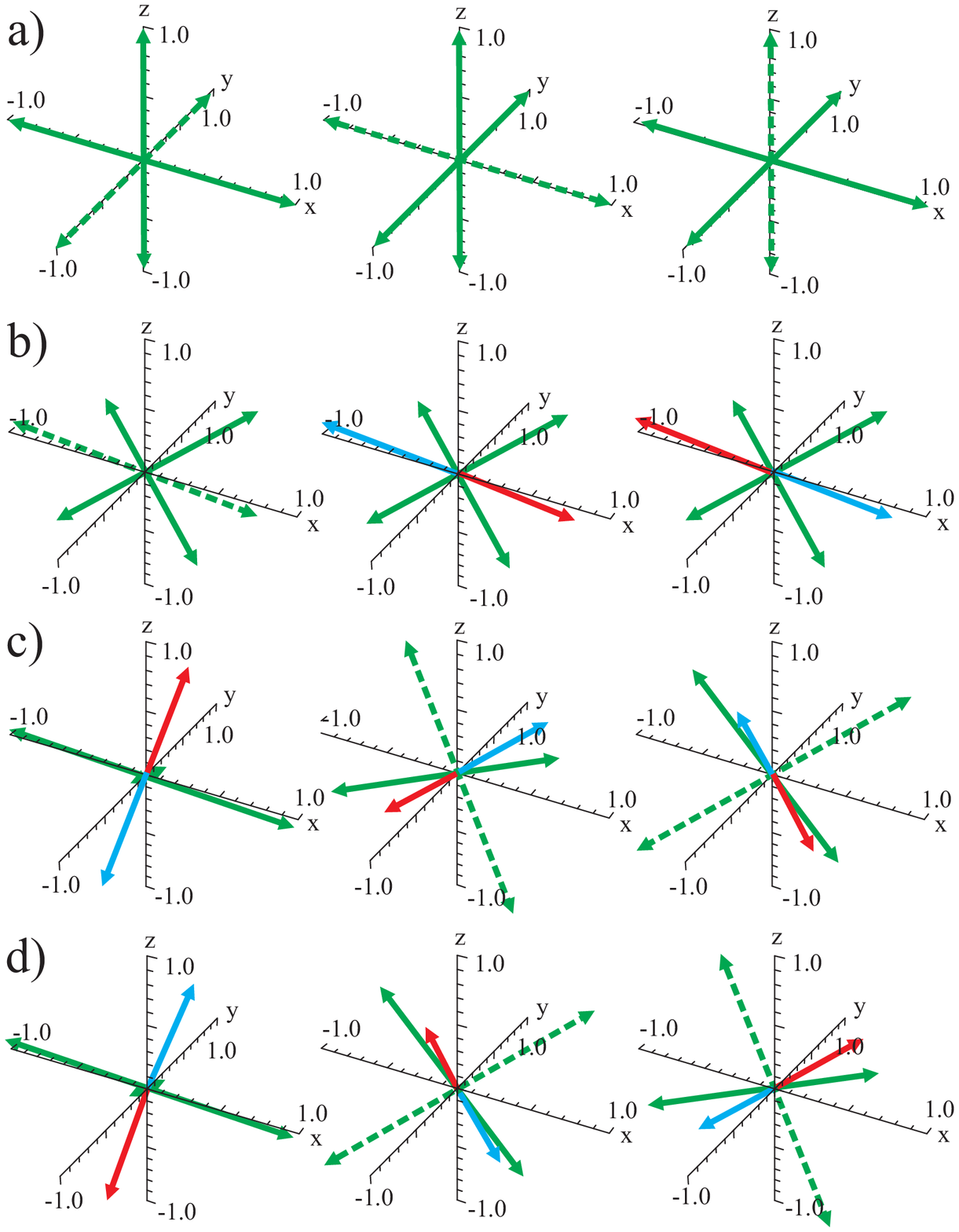}
        \caption{Representation of four qutrit MUBs -- Eqs. (\ref{MUB1}) -- a) states $|v_i^{(1)}\rangle$, b) states $|v_i^{(2)}\rangle$, c) states $|v_i^{(3)}\rangle$, d) states $|v_i^{(4)}\rangle$.}
        \label{fig5}
    \end{center}
\end{figure}

First, we present visualisation of mutually unbiased bases (MUBs) \cite{MUB}. We recall that two bases $\{|a_i\rangle\}_{i=1}^d$ and $\{|b_j\rangle\}_{j=1}^d$ in dimension $d$ are MUB if 
\begin{equation}
|\langle a_i | b_j \rangle| = \frac{1}{\sqrt{d}},~~\langle a_i | a_j \rangle = \langle b_i | b_j \rangle = \delta_{ij}.
\end{equation}  
In case of a qutrit there are four MUBs that can be represented as
\begin{eqnarray}
|v_1^{(1)}\rangle &=& \begin{pmatrix} 1 \\ 0 \\ 0 \end{pmatrix},|v_2^{(1)}\rangle = \begin{pmatrix} 0 \\ 1 \\ 0 \end{pmatrix},|v_3^{(1)}\rangle = \begin{pmatrix} 0 \\ 0 \\ 1 \end{pmatrix}; \label{MUB1} \\
|v_1^{(2)}\rangle &=& \frac{1}{\sqrt{3}}\begin{pmatrix} 1 \\ 1 \\ 1 \end{pmatrix},|v_2^{(2)}\rangle = \frac{1}{\sqrt{3}}\begin{pmatrix} 1 \\ \eta \\ \eta^{\ast} \end{pmatrix},|v_3^{(2)}\rangle = \frac{1}{\sqrt{3}}\begin{pmatrix} 1 \\ \eta^{\ast} \\ \eta \end{pmatrix}; \nonumber \\
|v_1^{(3)}\rangle &=& \frac{1}{\sqrt{3}}\begin{pmatrix} \eta \\ 1 \\ 1 \end{pmatrix},|v_2^{(3)}\rangle = \frac{1}{\sqrt{3}}\begin{pmatrix} 1 \\ \eta \\ 1 \end{pmatrix},|v_3^{(3)}\rangle = \frac{1}{\sqrt{3}}\begin{pmatrix} 1 \\ 1 \\ \eta \end{pmatrix}; \nonumber \\
|v_1^{(4)}\rangle &=& \frac{1}{\sqrt{3}}\begin{pmatrix} \eta^{\ast} \\ 1 \\ 1 \end{pmatrix},|v_2^{(4)}\rangle = \frac{1}{\sqrt{3}}\begin{pmatrix} 1 \\ \eta^{\ast} \\ 1 \end{pmatrix},|v_3^{(4)}\rangle = \frac{1}{\sqrt{3}}\begin{pmatrix} 1 \\ 1 \\ \eta^{\ast} \end{pmatrix}, \nonumber
\end{eqnarray}
where $\eta = \exp(\frac{2}{3}\pi i)$. We represent the above states in Fig. \ref{fig5}. Note, that in the spin 1 picture the basis states a) correspond to spin states $|s_j=0\rangle$, where $j$ corresponds to three mutually orthogonal directions. On the other hand, states b) correspond to eigenstates of a spin 1 operator $S_k$ along direction $k$ which is tetragonal to basis directions. The states c) and d) have a more complicated representation in terms of spin 1 operators. This problem was discussed in \cite{MUBspin1}. 

\subsection{Pseudo-qubit states}

Next, we consider an interesting class of qutrit states that resemble the whole class of qubit states, though in our case the states are never pure. We call them pseudo-qubit states. In case of single-qubit states the corresponding Bloch vectors lie on or inside a sphere. Here, we would like to mimic this property. We fix the metric tensor to be proportional to identity, which corresponds to fixing $\mathbf{\hat T} = \frac{1}{3} \openone$, and hence our ellipsoid becomes a sphere of radius $2/3$. The corresponding density matrix becomes 
\begin{equation}
\rho_{\mathbf a}=\begin{pmatrix}
\frac{1}{3} & \frac{-i a_z}{2} & \frac{i a_y}{2} \\
\frac{i a_z}{2} & \frac{1}{3} & \frac{-i a_x}{2} \\
\frac{-i a_y}{2} & \frac{i a_x}{2} & \frac{1}{3}
\end{pmatrix}.
\end{equation}
The matrix is uniquely determined by the three coordinates of a Bloch vector. It is non-negative if $\frac{4}{9} \geq {\mathbf a} \cdot {\mathbf a} \geq 0$. If ${\mathbf a} \cdot {\mathbf a}=\frac{4}{9}$ the state is of rank 2 and the corresponding eigenvalues are $\frac{2}{3}$, $\frac{1}{3}$ and $0$. Otherwise it is of rank 3. In case ${\mathbf a} \cdot {\mathbf a}=0$ the state is the maximally mixed qutrit state. These properties are analogous to the ones of qubits, the only difference is that the states are of different ranks.

The property,
\begin{equation}
\text{Tr}\{\rho_{\mathbf a} \rho_{\mathbf b}\} = \frac{1}{3} + \frac{{\mathbf a} \cdot {\mathbf b}}{2},
\end{equation}
resembles the one for qubits, although one cannot consider it as a scalar product or fidelity. The smallest possible value of ${\mathbf a} \cdot {\mathbf b}$ is $-\frac{4}{9}$ and hence the smallest value of 
$\text{Tr}\{\rho_{\mathbf a} \rho_{\mathbf b}\}$ is $\frac{1}{9}$.

\begin{figure}[t]
    \begin{center}
	\includegraphics[width=0.48\columnwidth]{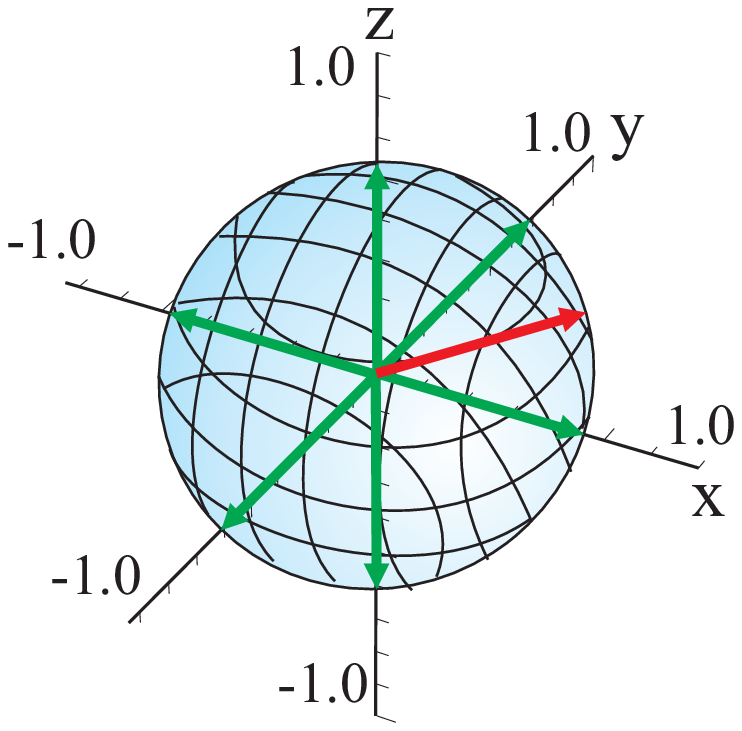}
        \caption{A visualisation of a rank 2 pseudo-qubit state.}
        \label{fig6}
    \end{center}
\end{figure}

Let us recall that a qutrit state can be considered as a symmetric state of two qubits. In this case the above class of pseudo-qubit states can be interesting from the point of view of studying the entanglement between the two qubits. If we use the positivity of the partial transpose as a criterion for separability, we observe that the maximally mixed qutrit state in the symmetric two-qubit representation is separable. In general, the pseudo-qubit state is separable for $ {\mathbf a} \cdot {\mathbf a} \leq \frac{1}{3}$. Interestingly, the state gets entangled by increasing a local parameter, however this feature should be carefully investigated in the future. 

%%%%%%%%%%%%%%%%%%%%%%%%%%%%%%%%%%%%%%%%%%%%%%%%%%%%%%%%%%%%%%%%%%%%

\section{Unitary dynamics} 

Unlike a qubit, the unitary dynamics of a qutrit cannot be represented only by rotations. In our representation such dynamics can also deform the ellipsoid and the vector $\mathbf{a}$. In general, the determinant of the density matrix is invariant under any unitary transformation which means that whatever we do with the ellipsoid and the vector, the quantity $ \mathbf{a}\cdot \mathbf{\hat\Gamma} \cdot \mathbf{a}$ must remain constant. The quantity $ \mathbf{a}\cdot \mathbf{\hat\Gamma} \cdot \mathbf{a}$ can be interpreted as a norm of the Bloch vector in a state dependent metric $\mathbf{\hat\Gamma}$ which is conserved by a unitary evolution. This is analogous to the fact that norm of the original qubit Bloch vector is also conserved.

\begin{figure}[t]
    \begin{center}
	\includegraphics[width=0.98\columnwidth]{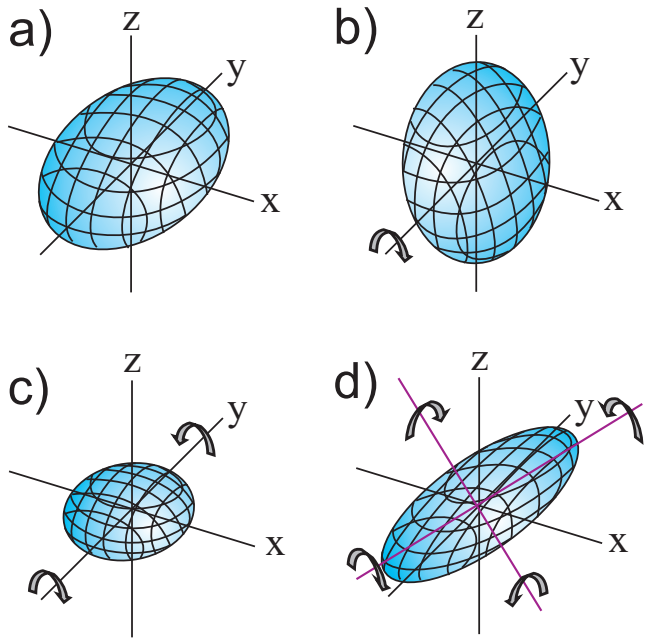}
        \caption{An ellipsoid affected by different types of unitary dynamics: a) original state; b) original state transformed by rotation generated by $S_y$; c) original state transformed by one-axis twisting generated by $S_y^2$; d) original state transformed by two-axis countertwisting generated by $A_y$.}
        \label{fig7}
    \end{center}
\end{figure}

Interestingly, the whole set of transformations can be intuitively represented by spin-1 squeezing \cite{Kitagawa}. In general there are three different types of spin-1 transformations. The first type is represented by rotations, which are generated by spin operators $S_j$. These transformations simply rotate our visualization about an axis $\mathbf{j}$ -- see Fig. \ref{fig7} a) and b).

The second type of transformation is known as \emph{one-axis twisting} and is generated by $S_j^2$. Using a slight simplification this transformation can be intuitively interpreted as a clockwise rotation about $\mathbf{j}$ of the part of the ellipsoid lying along $+j$ and an anti-clockwise rotation about the same direction of the remaining part of the ellipsoid lying along $-j$. Such a \emph{sqeezing} will obviously deform the ellipsoid and the vector $\mathbf{a}$ lying inside it. Moreover, it will change the corresponding principal semi-axes $\varepsilon_j$ and the metric tensor $\mathbf{\hat \Gamma}$ -- see Fig. \ref{fig7} a) and c). 

Finally, the third type of transformation is known as \emph{two-axes countertwisting} and is generated by $S_j^2 - S_k^2$, where the axes $\mathbf{j}$ and $\mathbf{k}$ are orthogonal. This operation can be interpreted as a composition of two one-axis twistings, the first along $\mathbf{j}$ and the second along $\mathbf{-k}$ -- see Fig. \ref{fig7} a) and d). Moreover, the operators $A_i = S_j S_k + S_k S_j$ introduced in the beginning generate such transformations. Note that $A_i = S_{i_+}^2 - S_{i_-}^2$, where $\mathbf{i_{\pm}}=(\mathbf{j}\pm\mathbf{k})/\sqrt{2}$.

%%%%%%%%%%%%%%%%%%%%%%%%%%%%%%%%%%%%%%%%%%%%%%%%%%%%%%%%%%%%%%%%%%%%

\section{Conclusions} 

We have introduced a three-dimensional representation of a qutrit in which the system is represented by a vector lying inside an ellipsoid. In this way the set of all qutrit states can be visualized in three dimensions as the set of all ellipsoids of semi-principal axes of length less or equal to $1$, each of which contains a vector lying inside or on the surface of the ellipsoid. We discussed the most important properties of this presentation, including characterization of pure and mixed states and the unitary evolution. The future investigations in this topic should include detailed descriptions of particular classes of qutrit states, like the pseudo-qubit states introduced above,  a detailed study of unitary and non-unitary dynamics, and a characterization of two-qutrit entanglement. The construction we provide can be possibly generalized for higher dimensional systems starting from fully symmetric states of many qubits.

%%%%%%%%%%%%%%%%%%%%%%%%%%%%%%%%%%%%%%%%%%%%%%%%%%%%%%%%%%%%%%%%%%%%

\emph{Acknowledgements.--} MM is supported by NCN Grant No. 2015/16/S/ST2/00447 within the project FUGA 4 for postdoctoral training. 
PK, WL and AK are supported by NCN Grant No. 2014/14/M/ST2/00818. PK was also supported by the National Research Foundation and Ministry of Education in Singapore.

\end{document}